\documentclass[%
 reprint,
 superscriptaddress,
 amssymb,
  aps,
 prl,
 longbibliography,
 floatfix,
]{revtex4-2}
\usepackage{graphicx, subcaption}
\usepackage{hyperref}
\usepackage{amsmath}
\usepackage{braket}
\usepackage{relsize}

\usepackage{booktabs}
\usepackage{xcolor}
\usepackage{soul}

\hypersetup{colorlinks,
    linkcolor={blue!75!black!80!yellow},
    citecolor={blue!75!black!80!yellow},
    urlcolor={blue!75!black!80!yellow}
    }

\begin{document}

\title{\large Low temperature decoherence dynamics in molecular spin systems using the Lindblad master equation }

\author{Timothy J. Krogmeier}
\affiliation{Department of Chemistry, Washington University in St. Louis, St. Louis, MO 61630 USA}
\affiliation{Department of Chemistry, University of Minnesota, Minneapolis, MN 55455 USA}

\author{Anthony W. Schlimgen}
\affiliation{Department of Chemistry, Washington University in St. Louis, St. Louis, MO 61630 USA}
\affiliation{Department of Chemistry, University of Minnesota, Minneapolis, MN 55455 USA}

\author{Kade Head-Marsden}
\affiliation{Department of Chemistry, Washington University in St. Louis, St. Louis, MO 61630 USA}
\affiliation{Department of Chemistry, University of Minnesota, Minneapolis, MN 55455 USA}
\email{khm@umn.edu}


\begin{abstract}

Understanding the spin dynamics in low-temperature settings is crucial to designing and optimizing molecular spin systems for use in emerging quantum technologies. At low temperatures, irreversible loss occurs due to ensemble dynamics facilitated by electronic-nuclear spin interactions. We develop a combined open quantum systems and electronic structure theory capable of predicting trends in relaxation rates in molecular spin ensembles. We use the Gorini-Kossakowski-Sudarshan-Lindblad master equation and explicitly include electronic structure information in the decoherence channels. We apply this theory to several molecular systems pertinent to contemporary quantum technologies. Our theory provides a framework to describe irreversible relaxation effects in molecular spin systems with applications in quantum information science, quantum sensing, molecular spintronics, and other spin systems dominated by spin-spin relaxation. 

\end{abstract}

\maketitle

\section{Introduction}

Molecular spin systems are centerpieces of a variety of emerging technologies including quantum computing, sensing, single-molecule magnets, and spintronics.~\cite{Colacio:2021a,Ruiz:2020a,Villain:2006a,Clerac:2012a,Gatteschi:2015a,Rondinelli:2023a, Sanvito:2011a,Birgitta:2020a,Bayliss:2022} In addition to technological applications, molecular spin systems are fundamental components of contemporary problems in experimental and theoretical chemistry across many fields. In cold chemistry, for example, the nuclear spin states of ultracold molecular collisions can influence the branching ratios of product channels.~\cite{Tscherbul:2023a,Ye:2009a,Ye:2017a,Hutzler:2020a} Other recent experiments characterize the spin dynamics in molecular systems by measuring the spin-lattice, $T_1$, and spin-spin, $T_2$, relaxation rates.~\cite{Freedman2021,Freedman:2022a,Freedman2017,Rondinelli:2024a,Hadt:2020a,Lijewski:2013a,Eaton:2006a,Sessoli:2016aa} In the context of quantum computing magnetic molecules could serve as quantum bits, or qubits, due to their optical addressability, synthetic tunability, and potential for high-temperature operation.~\cite{Richard2009,Carretta:2024a,Alexandrova:2021a,Wasielewski:2023a,Carretta:2024b} Fundamental understanding of the electron dynamics and the manipulation and control of the electronic state are essential for effective use of molecular spin systems in practical settings.

The dynamics, and therefore control, of these molecular systems are complicated by the ensemble nature of experiments performed in solution or solid phase. From a computational perspective, modeling ensembles requires an understanding of the distribution of configurations and how the individual molecular dynamics differ from one another in the ensemble. Additionally, experiments performed on single molecules still require detailed knowledge of these factors, as measurements of these single-molecule quantum systems must be performed many times to produce accurate statistics of the measurement outcomes.~\cite{deSousa:2009a} Furthermore modeling the $T_2$ relaxation, or dephasing, requires characterization of the connection between ensemble-averaged dynamics and loss of phase information. Modeling these relaxation processes is therefore a significant computational challenge because a Hilbert space containing every molecule in a solution is intractable. However, accurate modeling of ensemble dephasing would be a boon in informing experimental design and predicting $T_2$ relaxation times in molecular spin systems in contemporary technologies. 

In many technological applications, including molecular qubits and single molecule magnets, a two-level system is generated via an external magnetic field interacting with an unpaired electron. For practical use of a superposition of the parallel and antiparallel states, relative phase information must be long-lived. Dephasing occurs due to spin-lattice and spin-spin relaxation; at high temperatures, $T_1$ is expected to dominate, while at low temperatures $T_2$ dominates.~\cite{Freedman:2022a,Hadt:2020a,Lijewski:2013a,Eaton:2006a,Sessoli:2016aa}

In this \emph{Article}, we develop a phenomenological model capable of describing ensemble dynamics of molecular spin systems interacting with an environment. Because these systems undergo relaxation, one natural approach is to use an open quantum system perspective. We focus on the low-temperature setting using the Gorini-Kossakowski-Sudarshan-Lindblad (GKSL) equation to simulate the quantum dynamics.~\cite{Breuer2007,Gorini1976,Lindblad1976,Manzano2020} We incorporate the geometric and dynamic distributions present in ensemble systems to predict $T_2$ relaxation. \textcolor{black}{Specifically, we use \emph{ab initio} electronic structure to determine the GKSL decoherence rates for non-interacting ensembles of molecular spins.} We demonstrate the utility of this approach by simulating the decoherence dynamics of two molecular spin systems. This approach is general to cold molecular spin systems undergoing decoherence dynamics with applications in a wide array of emerging quantum technologies.

\section{Theory and Methods}

\begin{figure}[ht!]
    \centering
    \includegraphics[width = \columnwidth,trim=21cm 21cm 23cm 8cm,clip]{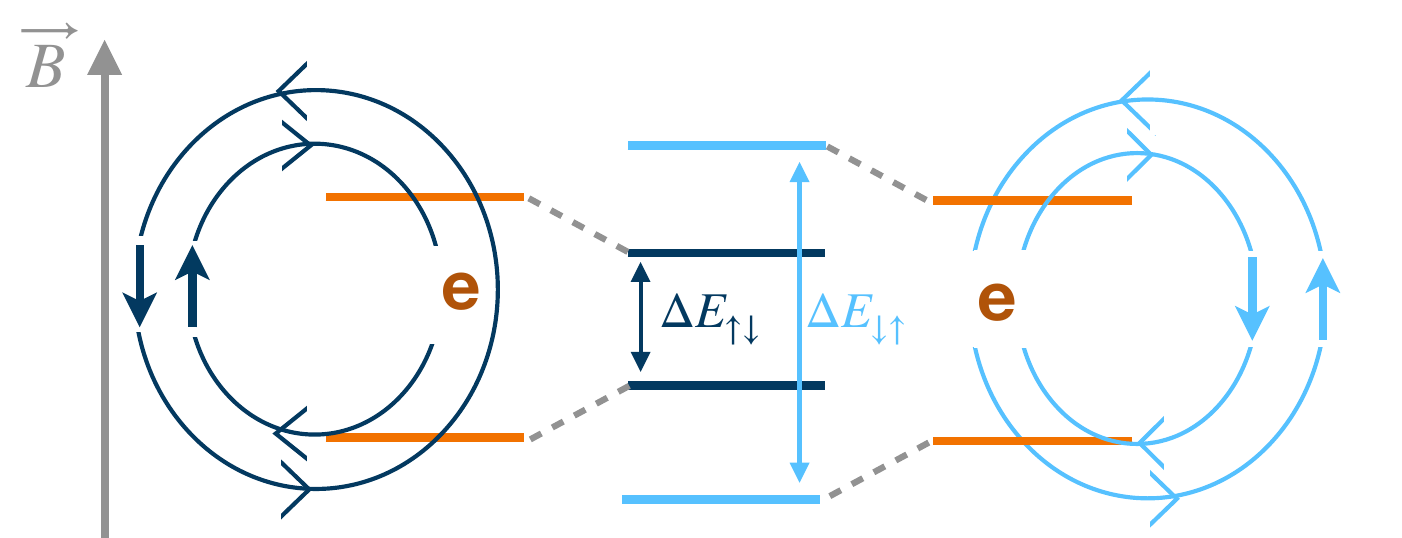}
    \caption{Two different pair orientations of nuclear spins, $\ket{\downarrow \uparrow}$ and $\ket{\uparrow \downarrow}$, interacting with an electron are shown on the left and right, respectively. Their different orientations and distances from the electron rescale the Zeeman energy, as shown in the middle of the diagram.}
    \label{fig:pair_orientations}
\end{figure}

One mechanism that leads to $T_2$ relaxation is nuclear spins coupled through a dipolar magnetic interaction, $\hat{H}_{dip}$, which connects the low-spin states of the two nuclei, and is proportional to $r^{-3}$ where $r$ is the internuclear distance.~\cite{Ramsey:1953a,Lazzeretti:2003a,Purcell:1952a} This coupling induces an exchange of spin orientation between two antiparallel nuclear spins in a pair-wise \emph{flip-flop} of magnetization, for example when the spin state changes from $\ket{\uparrow \downarrow}$ to $\ket{\downarrow \uparrow}$. In a molecular spin system a given spin state of the nuclear pair will induce a different magnetic field at the location of the electron, compared to the corresponding flip-flop state. Therefore, the flip-flop of a nuclear spin pair that are hyperfine coupled to the electron results in a time-dependent magnetic field at the electron, as shown in Figure~\ref{fig:pair_orientations}.
In either a time- or ensemble-averaged experiment, this time-dependent magnetic field rescales the Zeeman energy of the electron, broadening the energy states by dynamically changing the precession frequency of the electron. In a spin echo experiment at low temperatures, decoherence due to static inhomogeneities of an external magnetic field is removed, but the dynamic flip-flop broadening processes will remain. The time-dependent field induced by the nuclear spin flip-flops is therefore the most significant remaining contributor to dephasing.~\cite{Hahn1950,Sham2006,RenBao2012,Aruachan:2023} Modeling the ensemble spin-decoherence in these systems is therefore challenging due to the influence of the static electronic structure along with the irreversible nature of the electron broadening process. 

Here, we use the GKSL equation enhanced with \emph{ab initio} electronic structure to model the spin decoherence of an ensemble of molecules. \textcolor{black}{The GKSL equation is a commonly used open quantum system master equation derived under the Born-Markov approximation, meaning it can be used to capture the dissipative dynamics in quantum systems weakly coupled to quickly relaxing environments.~\cite{Gorini1976, Lindblad1976, Breuer2007, Manzano2020} This equation has found application in many fields including quantum biology, quantum information, quantum optics, and condensed matter physics.~\cite{Lidar1998, Plenio2008, Kraus2008, Gauger2011, Prosen2011, McCauley2020, Manzano2020} A major benefit of the GKSL equation is that, by construction, it preserves the positivity of the density matrix.~\cite{Gorini1976, Lindblad1976, Breuer2007, Manzano2020} This equation takes the form,}
\begin{equation}
     \frac{d\rho}{dt} = -\frac{i}{\hbar}\left[ \hat{H},\rho\right] + \mathlarger{\sum}_K\gamma_K(\hat{C}_K\rho \hat{C}_K^\dag - \frac{1}{2}\{\hat{C}_K^\dag \hat{C}_K,\rho\}),
     \label{eq:lindblad}
 \end{equation}
 where $\rho$ is the system density matrix, $\hat{H}$ is the system Hamiltonian, $\hat{C}_K$ are the decay channel operators with associated $\gamma_K$ decay rates, $\dag$ represents the adjoint operator, and $[\cdot,\cdot ]$ and $\{ \cdot,\cdot \}$ are the commutator and anticommutator, respectively. For each nuclear spin pair $i,j$ we compute the difference in magnetic environments.

\textcolor{black}{In typical ensemble experiments measurements will sample a distribution of molecular configurations or geometries. For non-interacting systems at low temperature, the distribution of geometries can be approximated from the zero-point vibrational modes of a molecule. To simulate such an ensemble, we sample the zero-point vibrational modes of each molecule using density functional theory (DFT), which results in a distribution of local magnetic environments for the nuclear spins. Computing this distribution is an important part of our approach because it estimates the broadening of the rescaled Zeeman energy in Fig.~\ref{fig:pair_orientations} which is the source of the irreversible decoherence in a molecular ensemble. For each molecule we generate $2\cdot(3N-6)$ different geometries from the normal modes.}

\textcolor{black}{We carry out all} electronic structure calculations using the ORCA 5 quantum chemistry software package.~\cite{Neese2020} Since the vanadium-oxo complexes are not strongly correlated, DFT is sufficient to describe their electronic structure. \textcolor{black}{We used t}he B3LYP hybrid functional, in addition to a triple-$\zeta$ polarized basis set for the carbon, hydrogen, oxygen, and sulfur, with a core-polarized CP(PPP) basis set for the vanadium ion, in addition to an auxiliary basis set.~\cite{Weigend2006,Becke:1988a,Parr:1988a} For each molecule \textbf{V1}-\textbf{V4} in Figure~\ref{fig:VO_series}, \textcolor{black}{we optimized} geometries optimized, and from those equilibrium geometries, \textcolor{black}{we calculated} vibrational modes calculated. \textcolor{black}{We followed t}he same process for \textbf{CuS} and \textbf{CuSe}, using the M06L functional with an ANO-RCC-DZP basis for the carbon and hydrogen, and an ANO-RCC-TZP basis for the copper, sulfur, and selenium, in addition to auxiliary basis sets.~\cite{Truhlar:2006a,Neese:2017a,Roos:1990a,Widmark:2004a,Widmark:2005a,Lenz:2017} The functionals and basis sets were chosen following previous work which qualitatively matches experimentally measured spin densities.~\cite{Freedman:2019a,Stanton2020}

During each calculation, \textcolor{black}{we computed} hyperfine tensors, taking into account dipole-dipole interactions, spin-orbit coupling, and the Fermi contact interaction between the electron and the spin-active hydrogen and vanadium nuclei in the molecules. \textcolor{black}{We computed the hyperfine tensors via an expectation value taken over the spin density of each molecule.}~\cite{Neese2020,Lazzeretti:2003a,Purcell:1952a,Ramsey:1953a} \textcolor{black}{We used the resulting hyperfine tensors to compute the difference in magnetic environment between two nuclear spins $i$ and $j$ as $\Delta_{ij}=|A_i-A_j|$, with full details for electronic structure methods and computation of hyperfine tensors shown in the Supplementary Information (SI).} 

\textcolor{black}{Following an analogous process for nuclear spin lattices with defects,~\cite{Sousa2003} we calculate the flip-flop rate} for each nuclear pair $i,j$ in a \textcolor{black}{specific geometry} according to, 
\begin{equation}
    T_{ij} = 2\sqrt{2\pi}\mathcal{A}(I)\frac{J_{ij}^2}{\kappa_{ij}}\mathrm{exp}\left(-\frac{\Delta^2}{8\kappa_{ij}^2} \right).
    \label{eq:ff_rate}
\end{equation}
$J_{ij}$ is the dipolar coupling strength between nuclei $i$ and $j$, given by,
\begin{equation}
    J_{ij} = -\frac{1}{4}\gamma_i\gamma_j\hbar\mu_0\frac{1 - 3\cos^2\theta_{ij}}{r_{ij}^3},
    \label{eq:dd_coupling}
\end{equation}
where $\gamma_i$ is the gyromagnetic ratio of nucleus $i$, $\hbar$ is the reduced Planck's constant, $\theta_{ij}$ is the angle between the external magnetic field and the vector connecting nuclei $i$ and $j$, and $r_{ij}$ is the distance between nuclei $i$ and $j$. $\mathcal{A}(I)$ is a normalization factor, and $\Delta_{ij} = |A_i-A_j|$ is the hyperfine energy difference between nuclei $i$ and $j$. $\kappa_{ij}$ is given by
\begin{equation}
    \kappa_{ij} = \frac{16}{3}I(I+1)\mathlarger{\sum}_{n\neq i,j}(J_{in}-J_{jn})^2,
    \label{eq:ff_rate_kappa}
\end{equation}
where the $n$ index runs over all the nuclei in the molecule that are not part of the current flip-flop pair $ij$.~\cite{Sousa2003} $I$ is the spin of the bath nuclei, which is $\frac{1}{2}$ in the present work. For this work, vanadium and copper are approximated as a spin $\frac{1}{2}$ nucleus. This is not expected to be a significant source of error for the present model, as the difference in hyperfine energy between the vanadium and copper metal ions and the distal hydrogens is so large that nuclear flip-flops between the vanadium and other spin-active nuclei are prevented, and therefore do not contribute to decoherence of the electron. 

This equation is derived by considering the lineshape of a nuclear pair in a nuclear magnetic resonance (NMR) experiment.~\cite{Sousa2003} $\kappa_{ij}$ represents broadening of the NMR lineshape of a nuclear pair because of the dipolar interactions between the nuclear flip-flop pair and all other spin-active nuclei. The broadening of the lineshape allows the flip-flop process to occur even when the Zeeman energy of each nuclei differs.~\cite{Blank:2016a,Wenckebach:2021a,parker:1970a,Sellschop:1995a,Weiss:1953a} The terms $A_i$ and $A_j$ represent the center of each feature in the NMR lineshape, or the chemical shift, of the two nuclei, and are taken from the hyperfine tensors calculated with the electronic structure methods described above. \textcolor{black}{Flip-flop rates were} generated by setting $\Delta$=$\Delta_0$=$0$ and $\Delta$=$\Delta_{ij}$=$|A_i-A_j|$ in Eq.~\ref{eq:ff_rate}. For each pair $ij$ the distribution of flip-flop rates \textcolor{black}{present in an ensemble solution} is given by the set $\{T_{ij}\}$, from which the standard deviation of flip-flop rates, $\sigma_{\{T_{ij}\}}$ for each pair in an ensemble is calculated. \textcolor{black}{Calculating the distribution of flip-flop rates and their use as the decay parameter for molecular spin decoherence is a major contribution of this work. This provides a relationship between \emph{ab initio} quantum chemistry and the rate of nuclear spin flip-flops in molecular spin systems.}

To account for the irreversible loss of coherence due to ensemble dephasing of the electron in an ensemble of molecular spin systems the GKSL equation, Eq.~\ref{eq:lindblad} was used. The standard deviation of the flip-flop rates described above, $\sigma_{\{T_{ij}\}}$ are used as the decay rates, $\gamma_K$, each of which correspond to a Lindbladian channel $\hat{C}_K$ which is chosen to be the Pauli-$Z$ operator, $\hat{S}_z$. The use of rates of transitions between states, such as the transition between flip-flop states, as the decay rates in the GKSL equation has been previously used.~\cite{Divincenzo:2013a} Each nuclear pair $ij$ in the molecule is assigned a loss channel and the associated decay rate corresponding to the standard deviation of the flip-flop rate of that nuclear pair.

To treat the Liouvillian part of the dynamics, some approximation is necessary. The vanadium-oxo molecular systems contain 13 spin-active nuclei, in addition to the electron, which amounts to a $2^{14}$ dimensional Hilbert space, while the copper systems contain 9 spin-active nuclei in addition to the electron, for a $2^{10}$ dimensional Hilbert space. To minimize this cost, the Hamiltonian is factorized into clusters of nuclear spins in their interactions with the central electron, following the master equation cluster-correlation expansion (ME-CCE).~\cite{Ren-Bao:2008a,Ren-Bao:2009a,Ren-bao:2012a,Yang:2020,Sham:2006a,Galli:2024a, Onizhuk:2024b} \textcolor{black}{A major contribution of this work is the physically-motivated derivation of the decay rates $\gamma_K$, where previous approaches as well as implementations of ME-CCE have used $\gamma_K$ as an adjustable parameter to match experiment.~\cite{Galli:2024a, Aruachan:2023} Here, no adjustable parameters are present, and qualitative experimental agreement is achieved.} For the purposes of this work, clusters of two nuclear spins are sufficient since the mechanism of dephasing being considered here is the pairwise flip-flop of two nuclear spins at a time.~\cite{Sham:2006a} Finally, the Hamiltonian parameters that were factorized using CCE were taken from the equilibrium geometry of each molecule calculated using DFT. \textcolor{black}{Additional details on the calculation of spin dynamics is included in the SI.} 

\section{Results}

\begin{figure}
    \centering
    \includegraphics[width=\columnwidth, trim = 0cm 2cm 28cm 0cm, clip]{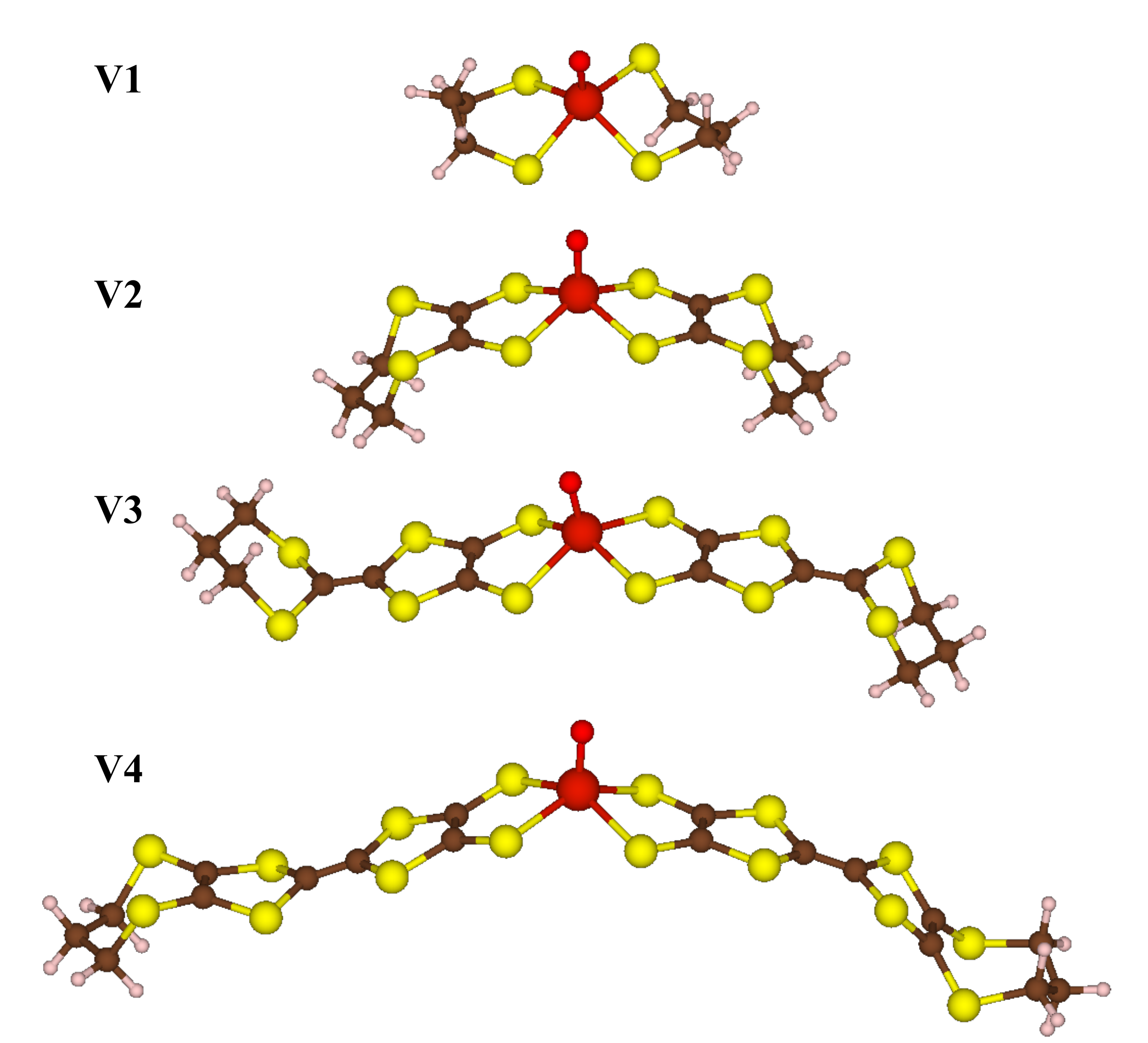}
    \caption{Series of vanadium-oxo molecular structures.}
    \label{fig:VO_series}
\end{figure}

The first system we consider is a series of vanadium-oxo structures \textbf{V1}-\textbf{V4} shown in Figure~\ref{fig:VO_series}. The $T_2$ spin-spin relaxation in these molecules has been experimentally measured, and the results highlight the counter-intuitive relationship between electron-nuclear distance and electron decoherence.~\cite{Freedman2017} The longest coherence was observed in the smallest molecule, \textbf{V1}, even though this molecule has the strongest hyperfine coupling interactions. These results were explained using the concept of the \emph{spin diffusion barrier}, a phenomenon in which nuclei close to a strong magnetic moment, in this case an electron, are energetically detuned relative to one another such that nuclear flip-flops are no longer energy conserving and become suppressed, thereby stifling decoherence.~\cite{bloembergen1949,blumberg1960,Khutsishvili1962,Khutsishvili1969,RORSCHACH196438,Monteiro2015,Stanton2020} In other words, for hyperfine tensors $A_i$ and $A_j$, when $\Delta_{ij}=|A_i-A_j|$ is very large, flip-flops and decoherence are both suppressed. If $A_i=A_j$ then $\Delta=\Delta_0=0$ and there is no suppression of flip-flops or decoherence due to the spin diffusion barrier.


\begin{figure}[ht!]
    \centering
    \includegraphics[width = \columnwidth]{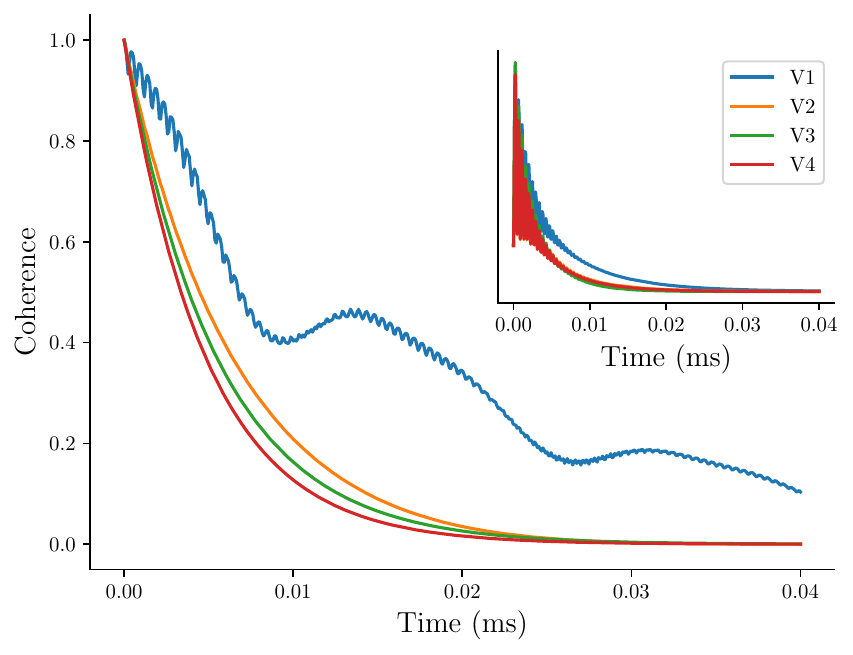}
    \caption{Simulated decoherence of the vanadium-oxo series using the GKSL equation, with experimental data reproduced from Ref.~\citenum{Freedman2017} shown in the inset.}
    \label{fig:exp_comp}
\end{figure}

In Figure~\ref{fig:exp_comp} we show the calculated coherence profiles using the GKSL equation for each ensemble of \textbf{V1}-\textbf{V4}, with the experimental data shown in the inset.~\cite{Freedman2017} \textcolor{black}{Experimental decoherence profiles are generally fit to an empirical equation (see SI) which provides an estimate of the decay constant. Following this procedure, Table~\ref{tab:decay_constants} shows the resulting decay constants which are fit from our simulated data.}
Our simulated results agree qualitatively with experiment, notably showing that \textbf{V1} has the longest-lived coherence, followed by \textbf{V2}-\textbf{V4} with faster decoherence. The simulations predict that \textbf{V2}-\textbf{V4} have similar decay constants, which is also in good agreement with the experiment.


\begin{table}[h!]
    \centering
     \begin{tabular}{@{\extracolsep{4pt}}l c c c} 
        \hline \hline
        \addlinespace[2pt]
         & $\Delta_{ij}$ & Exp. & $\Delta_0$     \\
        \cline{2-2}\cline{3-3}\cline{4-4}
        \addlinespace[2pt]
\textbf{V1}		&	0.0274	&  0.01011 &	0.00834\\
\textbf{V2}	&	0.0111	&  0.00602 &	0.00997\\
\textbf{V3}	&	0.0110	&  0.00659 &	0.00996\\
\textbf{V4}	&	0.00972	&  0.00595 &	0.00936\\
\hline\hline
    \end{tabular}
    \caption{Decay constants ($ms$) for the ensembles \textbf{V1}-\textbf{V4}, for the fits to simulated and experimental profiles. Details of the fitting function are shown in the SI.}
    \label{tab:decay_constants}
\end{table}

\begin{figure}[ht!]
    \centering
    \includegraphics[width = \columnwidth]{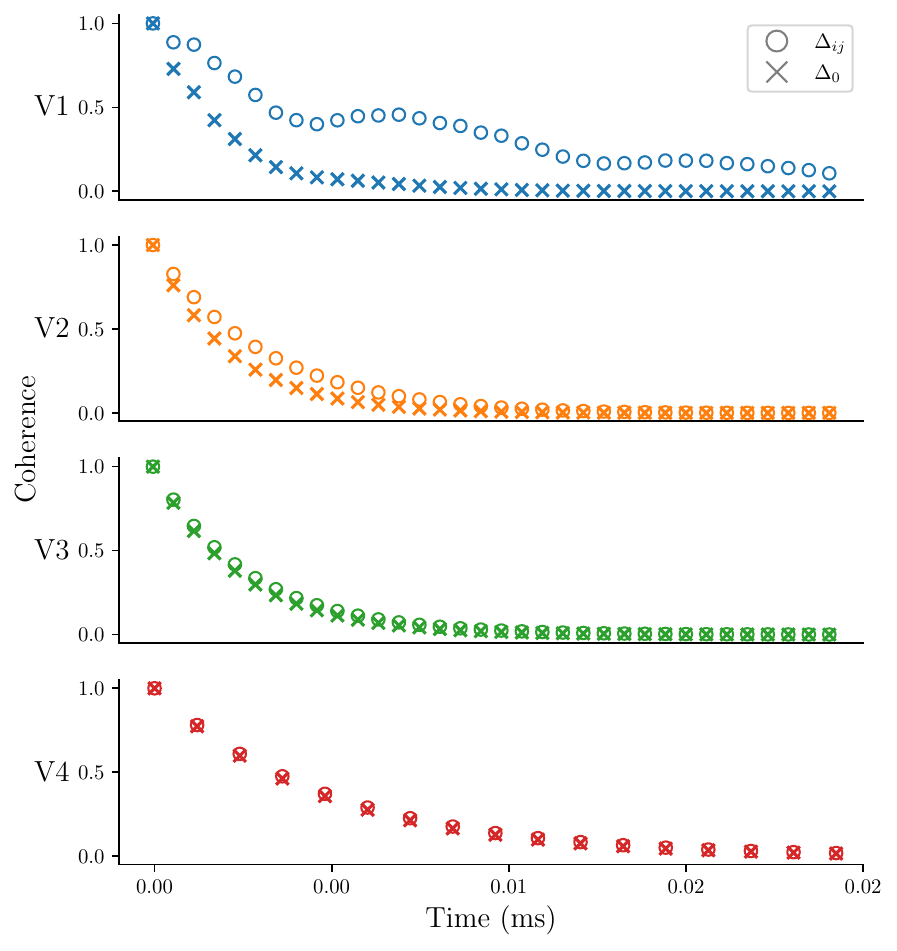}
    \caption{The coherence profiles for the four vanadium complexes using $\Delta_0$ and $\Delta_{ij}$.}
    \label{fig:trend_comp}
\end{figure}

To test the influence of the $\Delta$ parameter in Eq.~\ref{eq:ff_rate}, the dynamics are recalculated with $\Delta=\Delta_0$. The coherence profiles for \textbf{V1}-\textbf{V4} are shown in Figure~\ref{fig:trend_comp} comparing the coherence of each molecular ensemble with $\Delta_0$ and $\Delta_{ij}$. The decay time constants for the profiles with $\Delta_0$ are included in Table~\ref{tab:decay_constants}. 
Figure~\ref{fig:trend_comp} demonstrates that $\Delta_{ij}$ is crucial to reproduce the correct experimental trend. For $\Delta_0$, \textbf{V1} has the fastest decoherence time, whereas with $\Delta_{ij}$ it is longest-lived, which is what is observed in experiment. 

\begin{figure}[ht!]
    \centering
    \includegraphics[width = \columnwidth]{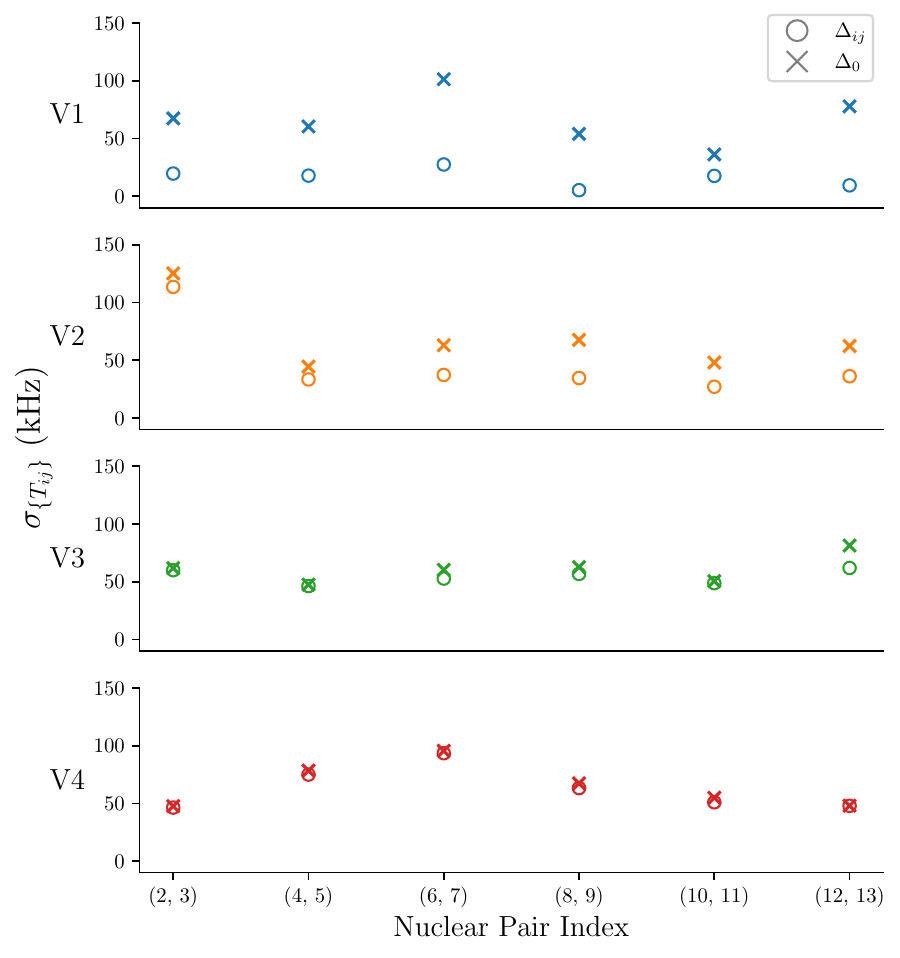}
    \caption{The standard deviations of the flip-flop rates for each ensemble \textbf{V1}-\textbf{V4} for $\Delta_0$ and $\Delta_{ij}$ are shown for the geminal hydrogens.}
    \label{fig:std_dev_comp}
\end{figure}

To further evaluate the impact of $\Delta$ on the spread of flip-flop rates, $\sigma_{\{T_{ij}\}}$ is plotted in Figure~\ref{fig:std_dev_comp} with $\Delta_{ij}$ and $\Delta_0$ for the nuclear pairs that have $\sigma_{\{T_{ij}\}}$ $>$ 5 kHz. \textcolor{black}{$\sigma_{\{T_{ij}\}}$ for all nuclear spins are shown in the SI.}
We show that \textbf{V1} has the greatest change in $\sigma_{\{T_{ij}\}}$ when changing from $\Delta_{ij}$ to $\Delta_0$, \textbf{V2} shows a slight change, followed by \textbf{V3}, and finally \textbf{V4} shows an even smaller change. This highlights the sensitivity of \textbf{V1}, and to a much lesser extent, \textbf{V2}-\textbf{V4}, to the $\Delta$ term. The $\Delta$ parameter encodes the hyperfine energy difference between each nuclei undergoing the mutual flip-flop process, and is therefore a measure of the different magnetic environments that each nuclei occupies.

\begin{figure}[ht!]
    \centering
    \begin{subfigure}[t]{0.45\linewidth}
        \centering
        \includegraphics[width=0.9\linewidth,trim=0cm 0cm 0cm 0cm,clip]{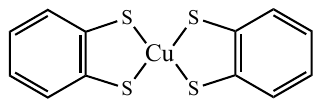}
        \caption*{\textbf{CuS}}
        \label{subfig:CuS}
    \end{subfigure}
    \begin{subfigure}[t]{0.45\linewidth}
        \centering
        \includegraphics[width=0.9\linewidth,trim=0cm 0cm 0cm 0cm,clip]{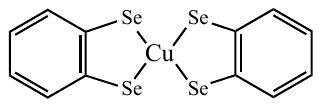}
        \caption*{\textbf{CuSe}}
        \label{subfig:CuSe}
    \end{subfigure}
    \caption{Two copper molecules, \textbf{CuS} and \textbf{CuSe}.}
    \label{fig:Cu_structures}
\end{figure}

The example of the vanadium ensembles demonstrates the importance of the location of the spin-active nuclei relative to the electron in determining decoherence. Altering the local magnetic environment of the electron can also be achieved by changing the ligand that is bonded to the metal center. To demonstrate this we compare the coherence profiles of a copper molecular spin system  with sulfur- or selenium-based ligands shown in Figure~\ref{fig:Cu_structures}, labelled as \textbf{CuS} and \textbf{CuSe}, respectively. Experiments show that \textbf{CuSe} has a longer-lived coherence due to increased copper-ligand covalency, indicating that the electron is delocalized on to the selenium ligand.~\cite{Freedman:2019a} \textcolor{black}{Computed spin densities for each molecule using the previously described DFT parameters are shown in the SI.}

\begin{figure}
    \centering
    \includegraphics[width = \columnwidth]{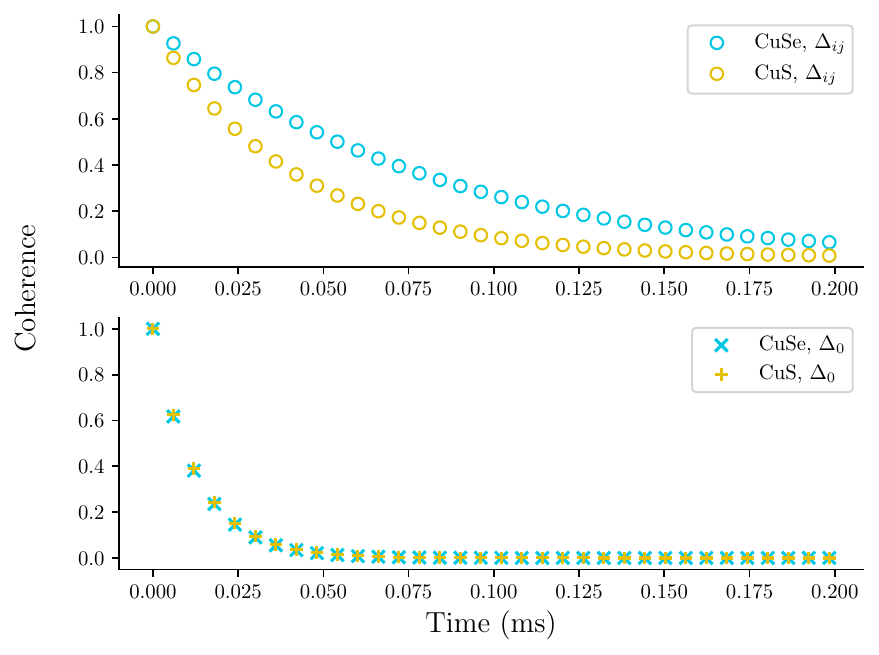}
    \caption{Coherence profiles for \textbf{CuS} and \textbf{CuSe} using $\Delta_{ij}$ (top) and $\Delta_0$ (bottom).}
    \label{fig:CuS_CuSe}
\end{figure}

Figure~\ref{fig:CuS_CuSe} shows simulated decoherence for these two molecular species using either $\Delta_{ij}$ or $\Delta_0$. The decay constants taken from fits to the calculated profiles are shown in Table~\ref{tab:CuS_CuSe_decay_consts}. 
Similar to the vanadium-oxo series, simulated decoherence trends are qualitatively comparable to experiments. Namely, the selenium based series has slower decoherence due to the delocalized copper-selenium $\sigma$-bond. This behavior is not seen when the local magnetic environment is neglected, i.e. when $\Delta=\Delta_0$. This highlights the importance of including the hyperfine energy difference in the calculation of the flip-flop rates for nuclear spin pairs. 


\begin{table}[h!]
    \centering
     \begin{tabular}{@{\extracolsep{4pt}}l c c c} 
        \hline \hline
        \addlinespace[2pt]
        & $\Delta_{ij}$ & Exp.    & $\Delta_0$    \\
        \cline{2-2}\cline{3-3}\cline{4-4}
        \addlinespace[2pt]
\textbf{CuS}	&	0.0411	&  0.002048&	0.0128	\\
\textbf{CuSe}	&	0.0766	&  0.003183&	0.0125\\
\hline\hline
    \end{tabular}
    \caption{Decay parameters ($ms$) for \textbf{CuS} and \textbf{CuSe}, along with experimental decay constants from Ref.~\citenum{Freedman:2019a}.}
    \label{tab:CuS_CuSe_decay_consts}
\end{table}


\section{Discussion and Conclusions}

Through these two \textcolor{black}{sets of} molecular spin systems, we have shown that the GKSL equation qualitatively reproduces experimental trends in spin decoherence. Importantly, the local magnetic environment of the decohering spin through electronic structure calculations \textcolor{black}{which directly relates to the decoherence rate of molecular spin ensembles}. The magnetic environment can be tuned in various ways, for example through increased metal-ligand covalency or through the internuclear distance of the spins. As an example of the former, we compare two copper molecules to show how increasing metal-ligand covalency can impact spin-spin decoherence. Increased metal-ligand covalency, or delocalization of the electron onto the selenium ligands, increases the hyperfine coupling via the Fermi contact and dipole-dipole interactions between the electron and the nuclear spins in the copper molecules. As an example of the latter, we compare a series of vanadium-oxo complexes showing the smallest complex exhibited the longest coherence time. Our results validate conventional understanding of decoherence in these molecular species; namely, that a spin diffusion barrier mitigates decoherence processes when nuclear spins are near the electronic spin. Our approach naturally incorporates the distinct magnetic environments of each nuclei by using \emph{ab initio} electronic structure in its treatment of the nuclear pair flip-flop process. In doing so, the presence of the spin diffusion barrier is a natural outcome, providing a theoretical justification for its use in explaining observed molecular $T_2$ trends. 

It is important to emphasize that the GKSL equation only includes the $T_2$ decoherence process, and does not include $T_1$ spin-lattice relaxation. The ensemble averaging over geometries via sampling of vibrational trajectories generates the distribution of flip-flop rates which yields the GKSL decay rate, but does not describe any coupling between the electron and lattice vibrations. Furthermore, the utility of the GKSL equation model in these systems is in describing trends, and informing experimental ligand choices for long-lived coherence times. While quantitative agreement with experiment is unlikely using this model, we obtain qualitative prediction of trends in coherence time with changes in electron-nuclear distance as well as atomic or ligand substitutions.

It is necessary for the master equation approach to encode the different magnetic environments of each nuclei in a spin pair to produce experimentally comparable results. This method explicitly incorporates the electronic structure of each molecule, and the distribution of flip-flop rates, into the decay rate in the GKSL equation. This approach is general to any molecular system at low temperature in a spin-spin dominated relaxation regime. This approach can be extended to strongly correlated transition metal systems, provided that the electronic structure can be computed accurately. The development of this genuinely open systems framework for molecular spin systems unlocks the possibility of modeling these systems with qualitative agreement with experimental results.

\section{Supplementary Information}
Supplementary Information includes the full standard deviation of the flip-flop rate for each nuclear pair in the ensembles \textbf{V1}-\textbf{V4}. The fitting function information used to extract decay times \textcolor{black}{from simulated data} is also provided. \textcolor{black}{Additional details on the electronic structure methods used and the ME-CCE method used are provided, including the computed spin densities of each molecule, calculation of hyperfine tensors, and the effective spin Hamiltonian used to calculate spin dynamics.}

\section{Author Contributions}

CRediT: Timothy J. Krogmeier:data curation, formal analysis, methodology, validation, visualization, writing-original draft, writing-review \& editing; Anthony W. Schlimgen: conceptualization, formal analysis, methodology, software, supervision, writing-original draft, writing-review \& editing; Kade Head-Marsden: conceptualization, formal analysis, funding acquisition, methodology, project administration, resources, software, supervision, writing-original draft, writing-review \& editing.

\section{Data Availability Statement}
Additional data and code that support the findings of this study are available from the corresponding author upon reasonable request \textcolor{black}{and on the Head-Marsden group's GitHub respository}.

\section{References}
\renewcommand*{\bibfont}{\normalsize}
\bibliography{main}

\end{document}